\def   \ni {\noindent}

\def   \ssk {\vskip  5truept}

\def   \bsk {\vskip 15truept}

\def   \newline {\hfil\break}

\documentstyle[epsfig]{article}
\begin{document}

\hsize 5truein
\vsize 8truein
\font\abstract=cmr8
\font\keywords=cmr8
\font\caption=cmr8
\font\references=cmr8
\font\text=cmr10
\font\affiliation=cmssi10
\font\author=cmss10
\font\mc=cmss8
\font\title=cmssbx10 scaled\magstep2
\font\alcit=cmti7 scaled\magstephalf
\font\alcin=cmr6 
\font\ita=cmti8
\font\mma=cmr8
\def\ref{\par\noindent\hangindent 15pt}
\null


\title{\ni Deep SIGMA observations of the central square degree of the Galaxy}                                               

\bsk \bsk
\author{\ni P.~Goldoni$^{1}$, A.~Goldwurm$^{1}$, F.~Lebrun$^{1}$, J.~Paul$^{1}$,
J.-P. Roques$^{2}$,E. Jourdain$^{2}$,L. Bouchet$^{2}$,P. Mandrou$^{2}$,E.Churazov$^{3,4}$,M. Gilfanov$^{3,4}$, R.
Sunyaev$^{3,4}$,A.Dyachkov$^{3}$, N. Khavenson$^{3}$,K. Sukhanov$^{3}$,I.Tserenin$^{3}$, N. Kuleshova$^{3}$}                                                       
\bsk
\affiliation{1) DAPNIA/Service d'Astrophysique, CEA/Saclay, F-91191 Gif-Sur-Yvette, France
\affiliation{2) Centre d'Etude Spatiale des Rayonnements, 9 Avenue du Colonel Roche, BP 4346, 31029 Toulouse Cedex France
\affiliation{3) Space Research Institute, Profsouznaya 84/32, Moscow 117810, Russia
\affiliation{4) Max-Planck-Institut f\"ur Astrophysik, Karl-Schwarzschild-Str. 1,85740 Garching
bei Munchen, Germany
}                                                
\bsk
\baselineskip = 12pt

\abstract{ABSTRACT \ni The Center of the Milky Way emits radiation through all the electromagnetic 
spectrum due to the presence of several astrophysical phenomena.
Various scientifical questions remain unsolved in the picture that emerge
from observations, the main one being the presence and activity of
a supermassive black hole at the Galactic Center.
Also in the X-ray band a strong diffuse emission is detected up to 22 keV,
it has been suggested that its origin is linked to past high energy
activity of the Galactic Center.
The French SIGMA soft gamma ray telescope on board the Russian Granat
satellite observed the Galactic Center (GC) for about 9 $\times$ 10$^6$ seconds from
1990 to the end of 1997. Its unique imaging capabilities coupled with the
unprecedented exposure already allowed us to set preliminary constraints
on the persistent high energy emission from Sgr A*.
Here we present a new analysis of the complete set of observations of this 
region taking into account the possible presence of diffuse emission
in the low energy channels. This allows us both to test current advection 
models of high energy emission from SgrA* and to investigate on past high
energy activity.
 }                                                    
\bsk
\baselineskip = 12pt

\bsk
\baselineskip = 12pt


\text{\ni 1. INTRODUCTION
\ssk
\ni     
\noindent The centre of our Galaxy, being the nearest Galactic Nucleus,
represents a fundamental object of study in modern astrophysics. However
conflicting conclusions have emerged from the analysis of observations in 
different wavebands. The most striking example of these problems
is the apparent paradox between the presence of a supermassive black hole
at the position of SgrA*, the source coincident with the dynamic centre of the Galaxy
(Eckart \& Genzel 1997) and its lack of activity in the X-ray domain (Goldwurm et al. 1994).

\noindent Recent advection models which produce accurate flux predictions
along all the electromagnetic spectrum have been proposed for Sgr A*
(Narayan et al. 1998). Flux and spectral distribution in the 1-100 keV band
are particularly critical in this model. However observations in these domains 
have long been limited by the poor angular resolution available at these
energies and by the source density in the GC region. A complex diffuse
emission is also present in the 3-20 keV band (Markevitch et al. 1993).
A careful subtraction of these contributions must be accomplished before 
evaluating SgrA* emission. In order to investigate these issues, we analyzed
all the hard X-ray observations of the central square degree of the Galaxy 
performed by the SIGMA telescope up to 1997.

\bsk
\ni 2. OBSERVATIONS AND RESULTS
\ssk
\ni 

\noindent The French hard X-rays (30-1300 keV) telescope SIGMA is
described in Paul et al. 1991. Launched on 1 December 1989 onboard the
Russian Granat Space observatory, SIGMA has observed twice per year up to 1997
the GC region for a total effective time of $\sim$ 9.2 $\times$
10$^6$ s obtaining hard X-ray images with 15' angular resolution
and typical 1$\sigma$ flux error of $<$ 2-3 mCrab (1 mCrab is about 8 $\times$
10$^{-12}$ erg cm$^{-2}$ s$^{-1}$ in the 40-80 keV band).

\noindent In the central square degree of the Galaxy SIGMA detected the X-ray 
burster A1742-294 and the unidentified source GRS 1743-290 (Goldwurm et al. 
1994) during several GC observations and the bursting pulsar
GROJ1744-28 in 1996 and 1997 (Bouchet et al. 1996). A1742-294 and GROJ1744-28 
are considered neutron star binaries while the nature of GRS 1743-290 is
less well determined.

\noindent GRS 1743-290 is a soft point source ($\alpha \sim$ 3) with a flux of
$\sim$ 10 mCrab in the 40-80 keV band. At lower (3-20 keV) energies the ART-P 
telescope detected extended emission centered near GRS 1743-290 best position (Markevitch et al. 1993)
but no point source coincident with the source's position (Pavlinsky et al. 1994). In this direction
several molecular clouds complexes are present (Morris \& Serabyn 1996) and the extended emission at
E$>$ 10 keV has been attributed to Thomson scattering on these clouds of
the flux from nearby compact sources in the present and in the past (Markevitch et al. 1993).
It has been deduced on the basis of this hypothesis that the 
central black hole did not emit at the Eddington luminosity for not even a
day in the last 400 years (Sunyaev et al. 1993). An extrapolation of the
ART-P spectrum of the extended emission with a power law with photon index 
$\alpha$=-2-3 implies a 40-80 keV flux in the range 5-20 mCrab depending on
the spectral index. These facts suggest that SIGMA could have detected
this emission along with the emission from the point source GRS 1743-290.

\smallskip
\noindent We produced light curves of GRS1743-290 in the 40-80 keV band for
all SIGMA observations. Our results indicate that the source is variable
GRS 1743-290 flux passes from 20 $\pm 4$ mCrab in Spring 1991 to less than 8 mCrab
(2 $\sigma$ upper limit) in Spring 1993. Its light curve is similar to
the one of the neutron star binary Terzan 2 (Vargas et al. 1997) in
this band. Detailed results will be presented elsewhere.

\noindent We also tried to fit an extended emission model to our images, our
model being a gaussian  with 40' FWHM convolved with the instrument's PSF,
extended along the galactic plane and centered on GRS 1743-290.
We performed our fit both to the total sum of SIGMA GC observations and to
the image of 1991 observations when 1E1740.7 -2942 had a very weak emission.
We made three different hypothesis on the nature of the emission at the position
of GRS 1743-290: a point source, an extended one and a sum of both. In this
case our results are preliminary but in the 1991 image unacceptable $\chi ^2$
were obtained with the diffuse emission model thus confirming the point source
nature of GRS 1743-290.

\noindent The case for the summed image is more complex as closer inspections
of SIGMA images show an irregular background in this region that could
seriously complicates the model of the source. In fact point
and diffuse source fits leave some extended structures at a level
of 2-3 standard deviations. The quality of the data thus makes
difficult to draw strong conclusions on the basis of our images.
Our result however exclude the presence of a strong diffuse source in the
1991 image and we can thus tentatively estimate an upper limit (1 $\sigma$)
of $\sim$ 10 mCrab (i.e. half of the flux of GRS 1743-290 on that date)
in the 40-80 keV band. This limit is compatible with an
extrapolation of the 3-20 keV emission detected by ART-P with a spectral
index $\alpha \sim$ -3.

\noindent We then searched for emission from SgrA*, after proper
subtraction of nearby sources. We did not detect
any emission, the results are summarized in Table 1, they are also shown
in Figure 1 along with soft X-rays upper limits defined following the
criteria of Narayan et al. (1998). A rough estimate
(case A) has been made using the sensitivity of our images to point sources
(not considering the influence of nearby sources). A more conservative
estimate (case B) has been made trying to fit along with the already
known sources an additional point source centered on SgrA* and calculating
which values are compatible with the observed shape.

\begin{figure}

\centerline{\psfig{figure=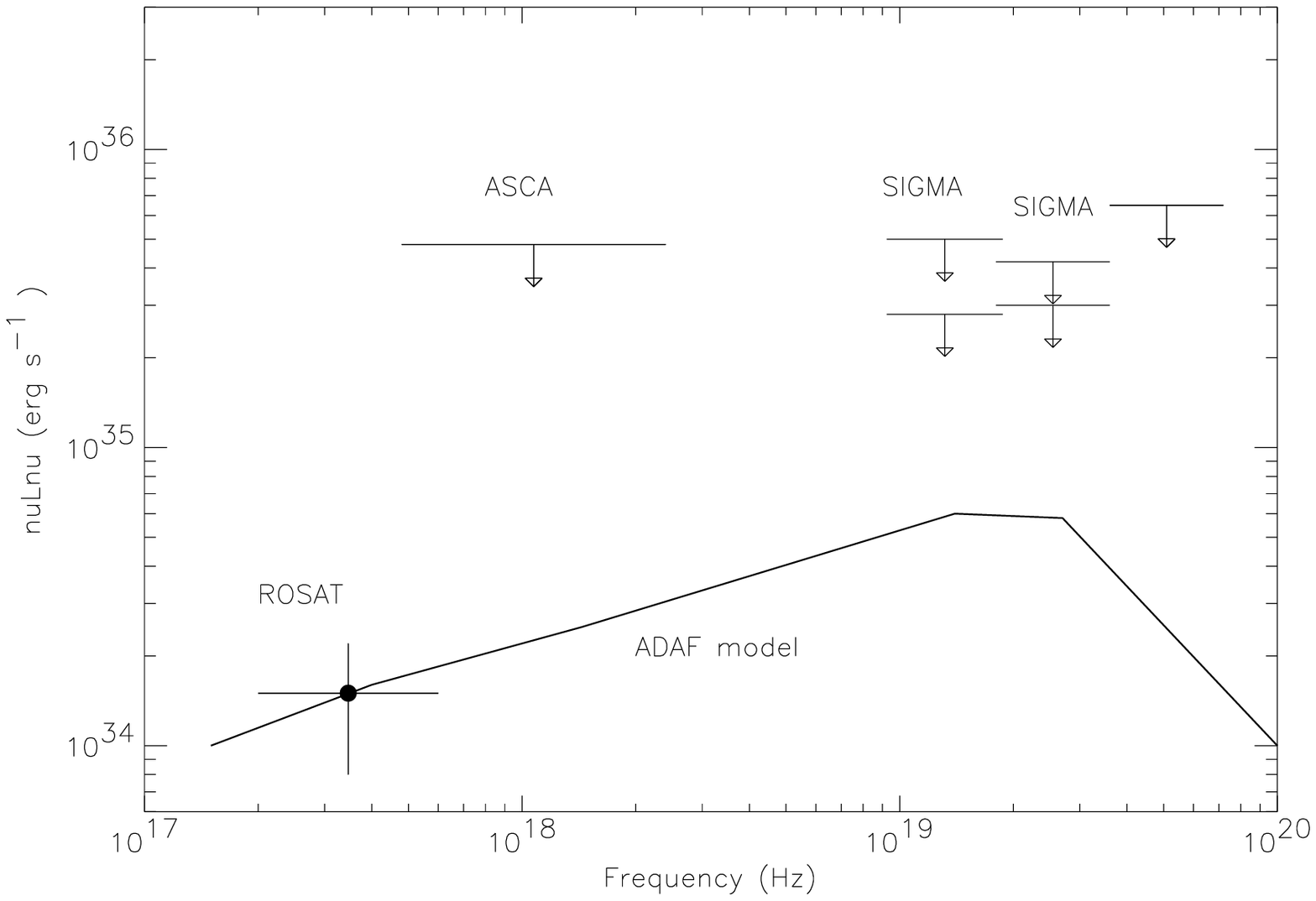,height=45mm,width=70mm}}

\caption {\bf Figure 1} {Soft X-rays and hard X-ray experimental data on the emission
of SgrA* are shown along with the ADAF model (Narayan et al. 1998). The soft X-ray points are taken
following the criteria of Narayan et al.}

\label{Figure1}
\end{figure}

\noindent At energies higher than 150 keV only 1E1740.7-2942 is present in the
field and only one upper limit is quoted. In general our limits are 50 \%
better than the ones presented by Goldwurm et al. (1994) and still no
emission is detected from Sgr A*.

\begin{table}
\caption{ \bf Table 1}{~~2$\sigma$ upper limits for SgrA* emission for a Crab-like
spectrum in cases A and B (see text for details).}
\label{Table1}
\[
    \begin{array}{ccc}
    \hline
\noalign{\smallskip}
${\rm Energy~range}$&${\rm Integrated~luminosity~(A)}$&${\rm Integrated~luminosity~(B)}$\\
${\rm keV}$ & ${\rm erg/s}$ & ${\rm erg/s}$ \\
\noalign{\smallskip}
\hline
\hline
\noalign{\smallskip}
40-80~${\rm keV}$  & 2.0 \times 10^{35} & 3.4 \times 10^{35}   \\
\hline
\noalign{\smallskip}
80-150~${\rm keV}$  & 2.0 \times 10^{35} & 2.8 \times 10^{35}  \\
\hline
\noalign{\smallskip}
150-300~${\rm keV}$  & 5.2 \times 10^{35} &  5.2 \times 10^{35}  \\
\hline
\noalign{\smallskip}

\noalign{\smallskip}
  \end{array} 
   \]
\end{table}



\bsk
\ni 3. CONCLUSIONS
\ssk
\ni 

\noindent We investigated on the nature of GRS1743-290 and we searched for hard
X-ray emission from SgrA*. We conclude that GRS1743-290 is probably
a point source, possibly a neutron star binary (Goldwurm et al. 1994)
Diffuse emission detected at lower energies has a flux lower than about
10 mCrab (1$\sigma$ statistical upper limit) in the SIGMA band.
We searched for emission from Sgr A* and we did not 
detect any signal obtaining improved upper limits. These limits are not 
constraining for present-day ADAF models and further observations in this
energy domain will be crucial for these models. In particular INTEGRAL Deep 
Galactic Center exposure will be able to detect the predicted flux
(Goldwurm et al. 1999).

\bsk
\baselineskip = 12pt
{\abstract \ni ACKNOWLEDGMENTS We acknowledge the paramount contribution of the SIGMA
Project Group of the CNES Toulouse Space Center to the overall success
of the mission. We thank the staffs of the Lavotchine Space Company, of
the Babakin Space Center, of the Baikonour Space Center, and the Evpatoria 
Ground Station for their unfailing support.

 }

\bsk
\baselineskip = 12pt


{\references \ni REFERENCES
\ssk
\ref  Bouchet L. et al., 1996, IAUC 6343
\ref  Eckart A. \& Genzel R. 1997 MNRAS 284, 576
\ref  Goldwurm A., Cordier B., Paul J., Ballet J., Bouchet L., et al., 1994, Nat, 371, 589
\ref  Goldwurm A., Goldoni P., Laurent Ph., Lebrun F., 1999, these Proceedings

\ref  Markevitch M., Sunyaev R.A. \& Pavlinsky M.N., 1993 Nature 364, 40
\ref  Morris M. \& Serabyn G., 1996, ARA\&A 34, 645
\ref  Narayan R. et al., 1998, ApJ 492, 554
\ref  Pavlinsky M.N., Grebenev S.A. \& Sunyaev R.A., 1994 ApJ 425, 110
\ref Paul J. et al. 1991, Adv. Space Res. 11 (8)289
\ref Sunyaev R.S., Markevitch M. \& Pavlinky M., 1993, ApJ 407, 606
\ref Vargas et al., 1997 Proc. 2nd INTEGRAL Symposium, eds. C. Winkler,
T. J.-L. Courvoisier, Ph. Durouchoux, ESA publications ESA-SP 382, p. 129

\end{document}